\documentclass[aps,pra,preprint,superscriptaddress,showpacs,nofootinbibfloatfix,amsmath,amsfonts,amssymb]{revtex4-2}%
\usepackage{amsmath,amsfonts,amssymb,color}
\usepackage{amsthm}
\usepackage{leftidx}
\usepackage{graphicx}
\usepackage{xcolor}
\usepackage{dcolumn}
\usepackage{bm}
\usepackage{epstopdf}
\usepackage{epsfig}
\usepackage{environ}
\usepackage{pdfcomment}

\usepackage{multirow}
\usepackage{setspace}
\usepackage{color}

\usepackage{float}
\usepackage[T1]{fontenc}
\usepackage[latin9]{inputenc}
\usepackage{setspace}
\usepackage{esint}

\begin{document}


\title{Entanglement phase transitions in non-Hermitian Kitaev chains}

\author{Longwen Zhou}
\email{zhoulw13@u.nus.edu}
\affiliation{%
	College of Physics and Optoelectronic Engineering, Ocean University of China, Qingdao, China 266100
}
\affiliation{%
	Key Laboratory of Optics and Optoelectronics, Qingdao, China 266100
}
\affiliation{%
	Engineering Research Center of Advanced Marine Physical Instruments and Equipment of MOE, Qingdao, China 266100
}

\date{\today}

\begin{abstract}
The intricate interplay between unitary evolution and projective measurements
could induce entanglement phase transitions in the nonequilibrium
dynamics of quantum many-particle systems. In this work, we uncover
loss-induced entanglement transitions in non-Hermitian topological
superconductors. In prototypical Kitaev chains with onsite particle
losses and varying hopping and pairing ranges, the bipartite entanglement
entropy of steady states is found to scale logarithmically versus
the system size in topologically nontrivial phases and become independent
of the system size in the trivial phase. Notably, the scaling coefficients
of log-law entangled phases are distinguishable when the underlying
system resides in different topological phases. Log-law to log-law
and log-law to area-law entanglement phase transitions are further
identified when the system switches between different topological
phases and goes from a topologically nontrivial to a trivial phase,
respectively. These findings not only establish the relationships
among spectral, topological and entanglement properties in a class
of non-Hermitian topological superconductors but also provide an
efficient means to dynamically reveal their distinctive topological
features.
\end{abstract}

\pacs{}
\keywords{}
\maketitle

\section{Introduction\label{sec:Int}}

The entanglement dynamics of open quantum many-body systems undergoing
nonunitary evolution have attracted significant attention in recent
years \cite{EPTRev01,EPTRev02,EPTRev03}. An intriguing phenomena
that could emerge in such contexts is the measurement-induced entanglement
transition \cite{EPT01,EPT02,EPT03,EPT04,EPT05,EPT06,EPT07}. It describes
a nonequilibrium phase transition of entanglement structures following
a quantum quench. In usual situations, measurement-induced entanglement
phase transitions originate from the competition between unitary
dynamics and quantum measurements. With the increase in measurement
rates, the bipartite entanglement entropy (EE) of nonequilibrium steady
states could undergo a transition from a volume-law to an area-law
scaling versus the system size. Despite great theoretical efforts
\cite{EPT08,EPT09,EPT10,EPT11,EPT12,EPT13,EPT14,EPT15,EPT16,EPT17,EPT18,EPT19,EPT20,EPT21,EPT22,EPT23,EPT24,EPT25,EPT26,EPT27,EPT28,EPT29,EPT30,EPT31,EPT32,EPT33,EPT34,EPT35,EPT36,EPT37,EPT38,EPT39,EPT40,EPT41},
measurement-induced entanglement transitions have also been explored
experimentally in setups including trapped ions and superconducting
qubits \cite{EPTExp01,EPTExp02,EPTExp03}, offering further insights
for understanding quantum information dynamics and simulating quantum
many-body~systems.

Open quantum systems described by non-Hermitian Hamiltonians constitute
an important context for exploring entanglement phase transitions.
Various types of non-Hermiticity-induced entanglement transitions
have been identified in gapped or critical non-Hermitian systems made
up of lattice fermions and quantum spin chains \cite{NHEPT01,NHEPT02,NHEPT03,NHEPT04,NHEPT05,NHEPT06,NHEPT07,NHEPT08,NHEPT09,NHEPT10,NHEPT11,NHEPT12,NHEPT13,NHEPT14,NHEPT15,NHEPT16,NHEPT17}.
In a non-Hermitian system with spatial nonreciprocity, the emergence
of non-Hermitian skin effects was found to accompany the transition
from a volume-law entangled to an area-law entangled phase in one
spatial dimension \cite{NHEPT05}. The development of a dissipation
gap in the energy spectrum of a non-Hermitian Hamiltonian was also
found to yield a volume-to-area law entanglement phase transition
in free-fermion chains \cite{NHEPT06}. Moreover, in non-Hermitian
systems with spatially uniform or quasiperiodic randomness \cite{NHEPT08,NHEPT09,NHEPT10,NHEPT11},
entanglement phase transitions beyond the conventional volume-law
to area-law scheme could emerge due to the interplay between disorder
and non-Hermitian effects. In addition, alternated and re-entrant entanglement
transitions may be engineered and controlled by time-periodic driving
fields in non-Hermitian Floquet systems \cite{NHEPT12}.

In this work, we continue the study of entanglement phase transitions
in non-Hermitian systems. We focus on one-dimensional (1D) topological
superconductors with onsite particle losses, which have been found
to possess rich topological phases and dynamical phase transitions
\cite{NHTSC01,NHTSC02,NHTSC03,NHTSC04,NHTSC05,NHTSC06,NHTSC07,NHTSC08,NHTSC09}.
In Section~\ref{sec:Mod}, we introduce our model and outline the approaches
of characterizing its spectrum, topological properties and entanglement
dynamics. In Section~\ref{sec:Res}, we explore entanglement phase transitions
in representative non-Hermitian Kitaev chains with varying ranges
of single-particle hopping and superconducting pairing terms. All-round
connections are established among the spectral structures, topological
transitions and entanglement phase transitions in the considered system.
Remarkably, each loss-induced transition between topologically distinct
superconducting phases is found to go hand-in-hand with a transition
in the scaling law of steady-state EE, i.e., an entanglement phase
transition. In Section~\ref{sec:Sum}, we summarize our results and discuss
potential future directions. Some further calculation details about
the non-Hermitian Hamiltonian and entanglement dynamics are provided in the Appendices \ref{sec:AppA} and \ref{sec:AppB}.

\section{Model and {Methods}
	\label{sec:Mod}}
We start by considering a generalized Kitaev chain \cite{KC} with
onsite particle losses, whose Hamiltonian can be expressed as

\begin{equation}
	\hat{H}=\frac{1}{2}\sum_{n}\left[\mu(2\hat{c}_{n}^{\dagger}\hat{c}_{n}-1)+\sum_{r}(J_{r}\hat{c}_{n}^{\dagger}\hat{c}_{n+r}+\Delta_{r}\hat{c}_{n}\hat{c}_{n+r}+{\rm H.c.})\right].\label{eq:H}
\end{equation}
Here, H.c. denotes Hermitian conjugation, $n\in\mathbb{Z}$ represents
the lattice index (with the lattice constant $a=1$), and $r\in\mathbb{Z}$
controls the ranges of hopping and pairing terms. $\hat{c}_{n}^{\dagger}$
($\hat{c}_{n}$) creates (annihilates) a spin-polarized fermion on
the $n$th lattice site. The chemical potential $\mu=u-iv$ is chosen
to be complex, where $u={\rm Re}\mu$, and $v=-{\rm Im}\mu>0$ describing
the loss rate. $J_{r}$ and $\Delta_{r}$ denote hopping and pairing
amplitudes of fermions over $r$ lattice sites {[}$r=1$ for the nearest-neighbor
(NN) coupling, $r=2$ for the next-nearest-neighbor (NNN) coupling,
etc.{]}. The presence of long-range hopping and pairing terms beyond nearest-neighbor ($r>1$)
allows the system to possess superconducting phases with large topological
invariants \cite{NHTSC07}. We will refer to the system described
by $\hat{H}$ in Equation~(\ref{eq:H}) with $v>0$ as the lossy Kitaev chain
(LKC). Note in passing that the effective non-Hermitian Hamiltonian in Equation~(\ref{eq:H}) can be
viewed as emerging from a stochastic evolution followed by post-selecting
null measurement outcomes. Some further derivation details of $\hat{H}$ are provided in
the Appendix \ref{sec:AppA}.

Taking the periodic boundary condition (PBC) $\hat{c}_{n}=\hat{c}_{n+L}$
and applying the Fourier transformation $\hat{c}_{n}=\frac{1}{\sqrt{L}}\sum_{k}e^{ikn}\hat{c}_{k}$,
we can express $\hat{H}$ in the momentum space as $\hat{H}=\frac{1}{2}\sum_{k}\hat{\Psi}_{k}^{\dagger}H(k)\hat{\Psi}_{k}$,
where $L$ is the length of lattice, $k\in[-\pi,\pi)$ is the quasimomentum, $\hat{\Psi}_{k}^{\dagger}=(\hat{c}_{k}^{\dagger},\hat{c}_{-k})$
is the Nambu spinor operator, and 
\begin{equation}
	H(k)=h_{y}(k)\sigma_{y}+h_{z}(k)\sigma_{z},\label{eq:Hk}
\end{equation}
\begin{equation}
	h_{y}(k)=\sum_{r}\Delta_{r}\sin(kr),\label{eq:hyk}
\end{equation}
\begin{equation}
	h_{z}(k)=\mu+\sum_{r}J_{r}\cos(kr).\label{eq:hzk}
\end{equation}
Here, $\sigma_{x,y,z}$ are Pauli matrices in their usual representations.
By diagonalizing $H(k)$, we can find the spectrum of the system as
$E_{\pm}(k)=\pm E(k)$, where
\begin{equation}
	E(k)=\sqrt{h_{y}^{2}(k)+h_{z}^{2}(k)}.\label{eq:Ek}
\end{equation}
With the loss rate $v>0$ in the complex chemical potential $\mu=u-iv$, $E(k)$
becomes complex in general. Meanwhile, $H(k)$ possesses the chiral
symmetry ${\cal S}=\sigma_{x}$, in the sense that ${\cal S}H(k){\cal S}=-H(k)$.
The topological phases of the system could then be characterized by
a winding number $w$ \cite{NHTSC07}, which is defined as

\begin{equation}
	w=\frac{1}{2\pi}\int_{-\pi}^{\pi}dk\partial_{k}\phi(k),\label{eq:w}
\end{equation}
where $\phi(k)=\arctan[h_{y}(k)/h_{z}(k)]$. The value of $2\pi w$
is equal to the accumulated winding angle $\phi(k)$ over the first
Brillouin zone (BZ) of $k$. As the imaginary part of $\phi(k)$ has no winding
\cite{DWNLee}, $w$ can only take real values. Moreover, when the
energy spectrum is gapped, $w$ will take an integer (a half-integer)
quantized value if the trajectory of vector $[h_{y}(k),h_{z}(k)]$
encircles an even (odd) number of exceptional points (EPs) of $H(k)$
following the change of $k$ from $-\pi$ to $\pi$ \cite{NHTSC07}.
The closing and reopening of spectrum gaps will also accompany the
quantized (or half quantized) changes of $w$. Therefore, we can characterize
the non-Hermitian topological phases and topological phase transitions
of our LKC by the winding number $w$ in Equation~(\ref{eq:w}).

In this work, we focus on the loss-induced entanglement phase transitions
and their connections with topological phases in non-Hermitian Kitaev
chains. To deal with the entanglement dynamics, we first prepare our
system in a certain initial state $|\Psi(0)\rangle$ at half-filling
and evolve it over a time duration $t$ according to the Hamiltonian
$\hat{H}$ {[}Equation~(\ref{eq:H}){]}. The resulting state of the system
then takes the form ($\hbar=1$)
\begin{equation}
	|\Psi(t)\rangle=\frac{e^{-i\hat{H}t}|\Psi(0)\rangle}{||e^{-i\hat{H}t}|\Psi(0)\rangle||},\label{eq:Psit}
\end{equation}
where the normalization factor $||e^{-i\hat{H}t}|\Psi(0)\rangle||$
could arise after taking the no-click limit of a monitored evolution \cite{NHEPT05,NHEPT06}.
With the normalized state $|\Psi(t)\rangle$, we could obtain the
single-particle correlator $C(t)$ of the system in position representation.
Due to the translational symmetry of our system, the matrix elements
$C_{m,n}(t)$ ($m,n=1,\ldots,L$) of $C(t)$ in real-space can be further
computed by performing the Fourier transformation of its related generator
in $k$-space, i.e.,
\begin{equation}
	C_{m,n}(t)=\frac{1}{L}\sum_{k}e^{ik(m-n)}C_{k}(t),\label{eq:Clmt}
\end{equation}
where the $2\times2$ matrix $C_{k}(t)$ can be obtained as
\begin{equation}
	C_{k}(t)=\frac{1}{2}\begin{pmatrix}1+\langle\sigma_{z}\rangle_{t} & \langle\sigma_{x}\rangle_{t}+i\langle\sigma_{y}\rangle_{t}\\
		\langle\sigma_{x}\rangle_{t}-i\langle\sigma_{y}\rangle_{t} & 1-\langle\sigma_{z}\rangle_{t}
	\end{pmatrix}.\label{eq:Ckt}
\end{equation}
The average $\langle\sigma_{j}\rangle_{t}$ ($j=x,y,z$) is taken
over the normalized state $|\psi_{k}(t)\rangle$, which is evolved by
$e^{-iH(k)t}$ from the initial state vector $|\psi_{k}(0)\rangle$
at each quasimomentum $k$ (see Appendix \ref{sec:AppB} for more
details). 

Finally, to obtain the bipartite EE, we decompose
our 1D chain of length $L$ into two spatially connected subsystems
A and B, whose number of lattice sites are $l$ and $L-l$, respectively.
Restricting the lattice indices $m$ and $n$ to the A segment gives
us the correlation matrix $C_{m,n}^{{\rm A}}(t)$ ($m,n=1,\ldots,l$)
of subsystem A. Then, according to the relation between single-particle
correlation matrix and bipartite EE of a Gaussian state \cite{PeschelEE},
we can express the EE between the two subsystems A and B as
\begin{equation}
	S(t)=-\sum_{j=1}^{l}[\zeta_{j}\ln\zeta_{j}+(1-\zeta_{j})\ln(1-\zeta_{j})],\label{eq:St}
\end{equation}
where $\{\zeta_{j}|j=1,\ldots,l\}$ are the eigenvalues of the $l\times l$
correlation matrix $C^{{\rm A}}(t)$. In numerical calculations, we
can obtain $S(t)$ efficiently at any time $t$ through Equations~(\ref{eq:Psit})--(\ref{eq:Ckt})
together with the diagonalization of $C^{{\rm A}}(t)$. In the long-time
limit, $S(t)$ will reach a stationary value $S(L,l)\equiv\lim_{t\rightarrow\infty}S(t)$,
which may depend on the total system size, the subsystem size and
other system parameters. We can numerically find the bipartite EE
$S(L,l)$ of steady states by considering an evolution time duration
$t\in[0,T]$ that is long enough, so that any variations of $S(t)$
over time are negligible for $t\geq T$.

In the next section, we explore entanglement phase transitions in
two representative LKC models with distinct topological properties.
We will see clear changes in the scaling behaviors of steady-state EE when
the considered system undergoes topological phase transitions. Moreover,
qualitatively different scaling laws of EE in steady states will be uncovered
when the system resides in topological and trivial non-Hermitian superconducting
phases, respectively.

\section{Results\label{sec:Res}}
In this section, we investigate entanglement phase transitions in
two non-Hermitian Kitaev chains with different ranges of hopping and
pairing terms. We first reveal the spectrum and topological properties
of each model. This is followed by the demonstration of their entanglement
dynamics and the related scaling laws of steady-state EE. Finally,
we establish the entanglement phase diagrams and further unveil their
connections with the topological phases and transitions for both models. 

\subsection{LKC with Nearest-Neighbor Hopping and Pairing\label{subsec:NNLKC}}

We start with a ``minimal'' model of LKC by restricting the hopping
and pairing terms to NN sites. Referring to Equation (\ref{eq:H}),
we choose $r=1$, let $(J_{1},\Delta_{1})=(J,\Delta)$, and set the
chemical potential $\mu=u-iv$. The resulting system is described
by the Hamiltonian
\begin{equation}
	\hat{H}_{1}=\frac{1}{2}\sum_{n}[\mu(\hat{c}_{n}^{\dagger}\hat{c}_{n}-1/2)+J\hat{c}_{n}^{\dagger}\hat{c}_{n+1}+\Delta\hat{c}_{n}\hat{c}_{n+1}+{\rm H.c.}].\label{eq:H1}
\end{equation}
It has been identified that under PBC, the two energy bands $E_{\pm}(k)=\pm E(k)$
{{[}}Equation~(\ref{eq:Ek}){{]}} of this NN LKC model become gapless at $E=0$
when the system parameters satisfy the condition $u^{2}/J^{2}+v^{2}/\Delta^{2}=1$
\cite{NHTSC07}. When $u^{2}/J^{2}+v^{2}/\Delta^{2}<1$, there is
a line gap lying along the ${\rm Re}E$ axis between the two bands
$\pm E(k)$ on the complex energy plane. Within the gap, a pair of
degenerate Majorana zero modes can be found at the edges of the chain
under the open boundary condition (OBC), and the system is topologically
nontrivial in this case~\cite{NHTSC07}. When $u^{2}/J^{2}+v^{2}/\Delta^{2}>1$,
a line gap in the energy spectrum is opened along the ${\rm Im}E$
axis, and there are no localized Majorana modes with zero energy at
system edges under the OBC. In this case, the system becomes topologically
trivial and possesses a trivial dissipation gap \cite{NHTSC07}.

\begin{figure}
	\includegraphics[scale=0.6]{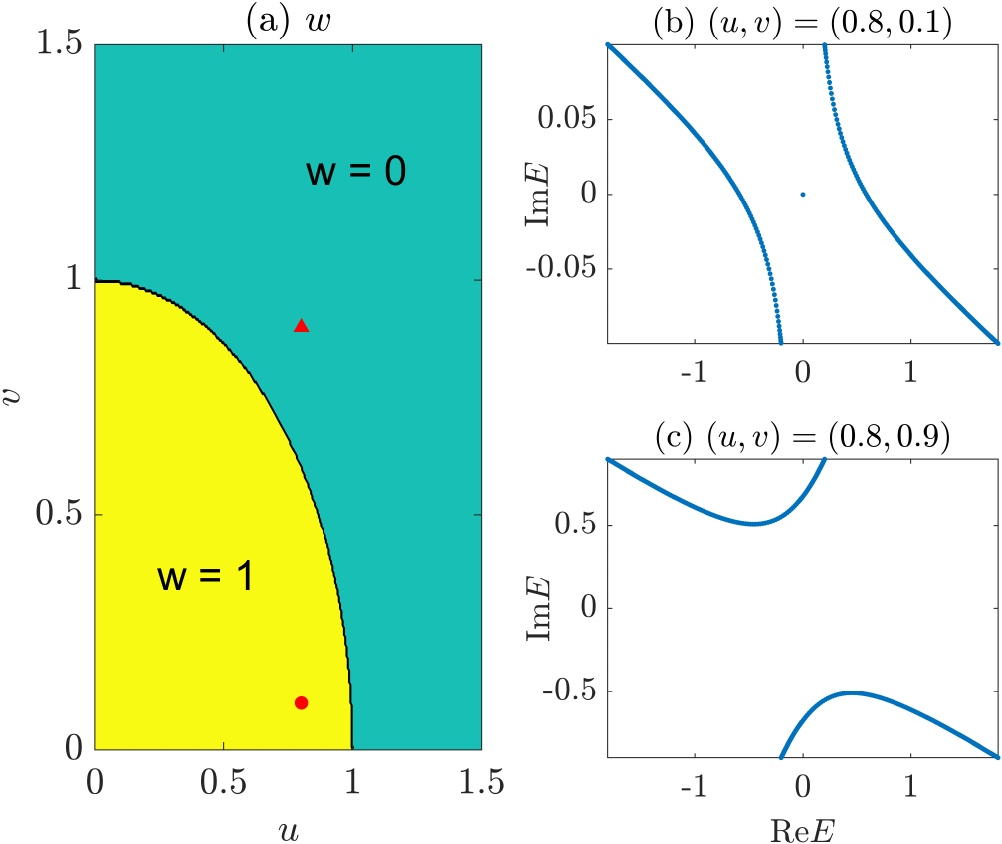}
	
	\caption{{Topological} 
		phase diagram and typical spectra of the LKC with NN hopping
		and pairing. (\textbf{a}) shows the winding number $w$ vs. the real and imaginary
		parts of chemical potential $u$ and $v$. The yellow and green regions
		have $w=1$ and $w=0$, respectively. The red solid dot in (\textbf{a}) resides
		at $(u,v)=(0.8,0.1)$, with the associated spectrum of $\hat{H}_{1}$
		shown in (\textbf{b}). The red solid triangle of (\textbf{a}) is located at $(u,v)=(0.8,0.9)$,
		and the associated spectrum of $\hat{H}_{1}$ is given in (\textbf{c}) on the
		complex energy plane. Other system parameters are $J=\Delta=1$ for
		all panels. \label{fig:LKC1EW}}
\end{figure}

In Figure~\ref{fig:LKC1EW}, we present typical energy spectra on the
complex plane and the topological phase diagram of the NN LKC model
{[}Equation~(\ref{eq:H1}){]}. In Figure~\ref{fig:LKC1EW}a, the phase
diagram is obtained by computing the topological winding number $w$
{[}Equation~(\ref{eq:w}){]} at different system parameters $(u,v)$. The
results confirm that the NN LKC indeed belongs to the topologically
nontrivial (trivial) phase with $w=1$ ($w=0$) when $u^{2}/J^{2}+v^{2}/\Delta^{2}<1$
($>1$), with the phase boundary $u^{2}/J^{2}+v^{2}/\Delta^{2}=1$
given by the black solid line between the yellow and green regions
in Figure~\ref{fig:LKC1EW}a. In Figure~\ref{fig:LKC1EW}b,c,
we illustrate the complex spectrum of the system in the topological
and trivial phases, respectively, under the OBC. A real energy gap
carrying a pair of Majorana edge modes at $E=0$ is observed in the
topological phase, while an imaginary energy gap with no Majorana
zero modes is found in the trivial phase, which validates our theoretical
predictions \cite{NHTSC07}.

\begin{figure}
	
	\includegraphics[scale=0.6]{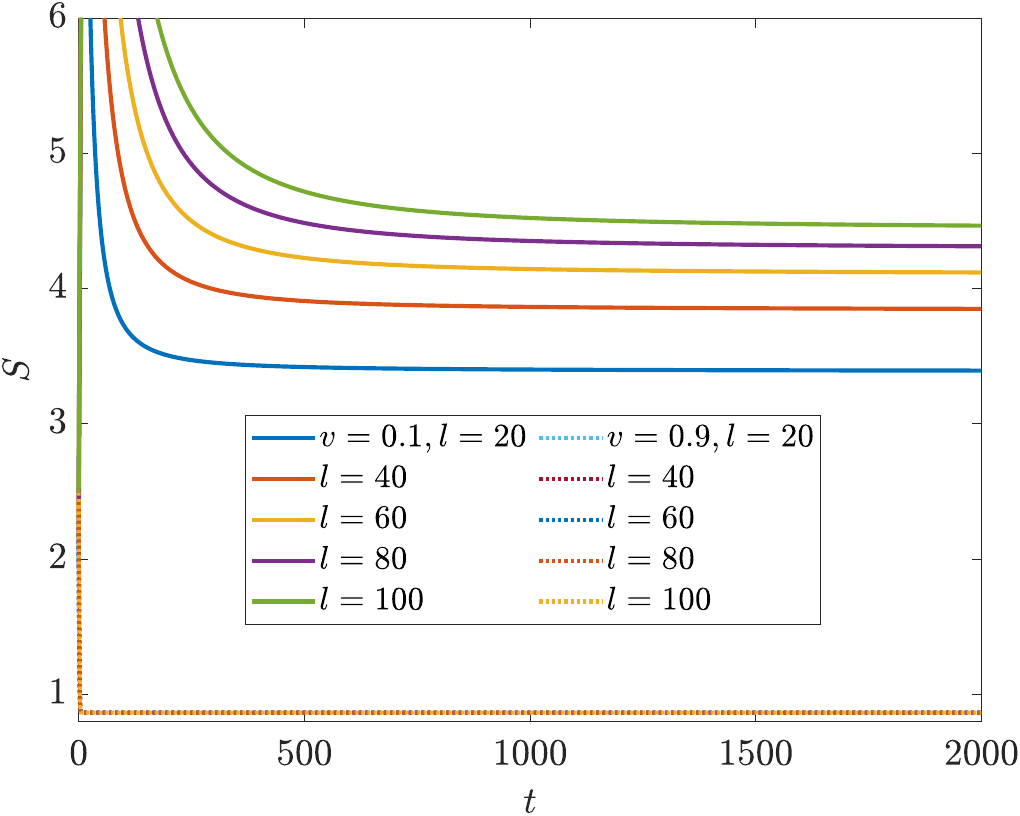}
	
	\caption{Bipartite EE vs. time $t$ for the LKC with NN hopping and pairing
		for the loss rate $v=0.1$ (solid lines) and $0.9$ (dotted lines).
		Other system parameters are $J=\Delta=1$ and $u=0.8$. $l$ denotes
		the subsystem size and the total lattice size is $L=2\times10^{4}$.
		\label{fig:LKC1EEvst}}
\end{figure}

{With} the above knowledge on the spectral and topological properties
of the NN LKC, it would be interesting to check whether the dynamics
and scaling laws of its EE could show distinct behaviors in different
topological phases. Following Equations~(\ref{eq:Psit})--(\ref{eq:St}),
we compute the evolution of bipartite EE $S(t)$ of the NN LKC under
PBC with a large system size $L$. A set of representative results
versus different sizes $l$ of the subsystem A and at different loss
rates are shown in Figure~\ref{fig:LKC1EEvst}. In all of the calculations,
we choose the initial state at different quasimomenta as 
\begin{equation}
	|\psi_{k}(0)\rangle=\frac{1}{\sqrt{2}}\begin{pmatrix}1\\
		e^{ik/2}
	\end{pmatrix},\label{eq:psik0}
\end{equation}
where $k=-\pi,-\pi+2\pi/L,\ldots,\pi-2\pi/L$. Other forms of pure and
non-stationary initial states generate consistent results regarding
the entanglement dynamics. We find that when the system parameters
satisfy $u^{2}/J^{2}+v^{2}/\Delta^{2}<1$, the bipartite EE will first
experience a transient time window with a non-monotonous behavior
in time, and finally evolving to a stationary value in the late time
regime. After the steady state is reached, the EE raises monotonically
with the increase in subsystem size $l$ (solid lines in Figure~\ref{fig:LKC1EEvst}).
On the contrary, in the parameter regime with $u^{2}/J^{2}+v^{2}/\Delta^{2}>1$,
the bipartite EE will converge to the same steady-state value for
any subsystem size $l$ after a transient time window, which implies
that $\lim_{t\rightarrow\infty}S(t)\sim l^{0}$ in this region (dashed
lines in Figure~\ref{fig:LKC1EEvst}). In summary, these observations
suggest that in the NN LKC, the scaling laws of steady-state EE vs.
the subsystem size could indeed be qualitatively different in the
topological phase with a real energy gap and the trivial phase with
an imaginary energy gap. A loss-driven entanglement phase transition
may then occur and go hand-in-hand with the topological phase transition
of the system.

\begin{figure}
	
	\includegraphics[scale=0.5]{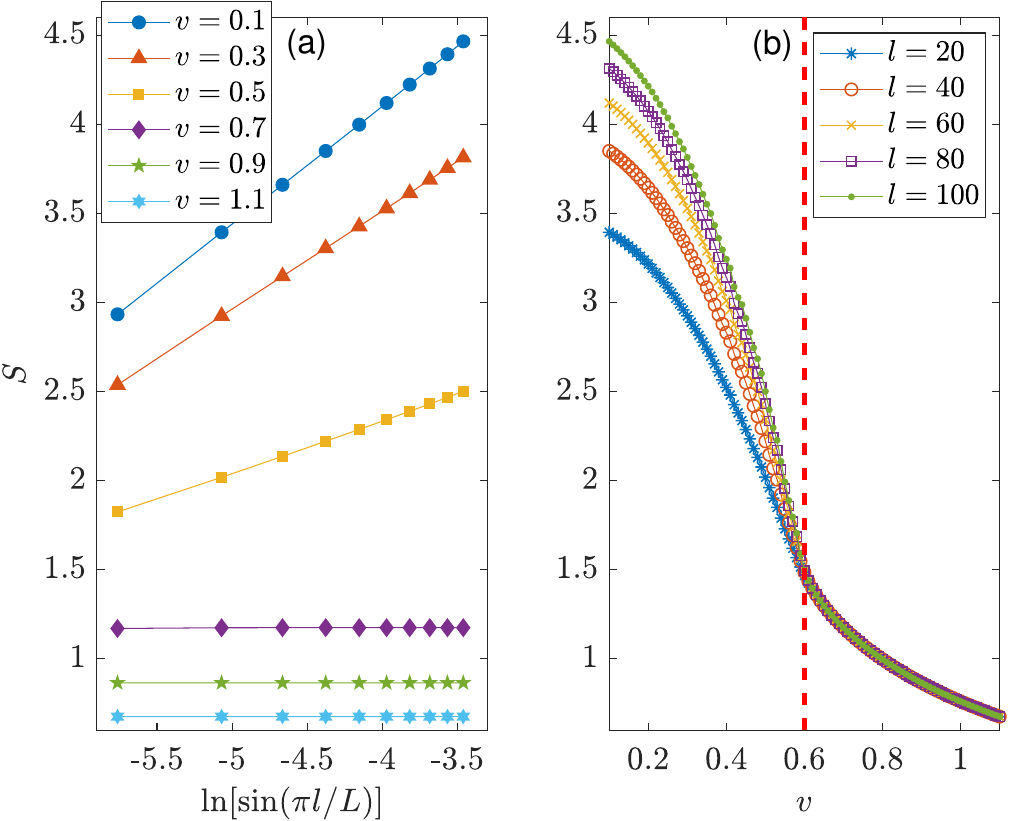}
	
	\caption{{Bipartite}  EE of steady states vs. (\textbf{a}) the subsystem size $l$, and
		(\textbf{b}) the loss rate $v$ for the LKC with NN hopping and pairing. Other
		system parameters are $J=\Delta=1$ and $u=0.8$ for both panels.
		The lattice size of the whole system is $L=2\times10^{4}$. The vertical
		dashed line in (\textbf{b}) highlights the phase transition point at $v=0.6$.
		\label{fig:LKC1EEvslv}}
\end{figure}

To further confirm the existence of different entangling phases in
the NN LKC, we present in Figure~\ref{fig:LKC1EEvslv} the steady-state
EE $S(L,l)$ versus the subsystem size $l$ and loss rate $v$. The
steady-state is reached by evolving the initial state $|\Psi(0)\rangle$
according to Equation~(\ref{eq:Psit}) {[}with $\hat{H}=\hat{H}_{1}$ in
Equation~(\ref{eq:H1}){]} over a long time duration, which is set to $T=2000$
in our calculations. In Figure~\ref{fig:LKC1EEvslv}a, we identify
the scaling-law of steady-state EE as $S(L,l)\sim\ln[\sin(\pi l/L)]$
for $u^{2}/J^{2}+v^{2}/\Delta^{2}<1$ and $S(L,l)\sim l^{0}$ for
$u^{2}/J^{2}+v^{2}/\Delta^{2}>1$, respectively, with $v\neq0$. In
the thermodynamic limit ($L\rightarrow\infty$), we obtain the following
scaling laws for the bipartite EE of steady states at half-filling,
i.e.,
\begin{equation}
	S(L,l)\sim\begin{cases}
		\ln l, & v\in(|\Delta|\sqrt{1-u^{2}/J^{2}},\infty),\\
		l^{0}, & v\in(0,|\Delta|\sqrt{1-u^{2}/J^{2}}).
	\end{cases}\label{eq:EE1vsl}
\end{equation}
Therefore, the EE satisfies a log-law (an area-law) vs. the subsystem
size in the weak (strong) dissipation regime of the system. These
two distinct entangling phases are clearly illustrated in the regions
with $v<0.6$ and $v>0.6$ in Figure~\ref{fig:LKC1EEvslv}b. Remarkably,
the separation point {[}red dashed line at $v=0.6$ in Figure~\ref{fig:LKC1EEvslv}b{]}
between these two entangling phases is precisely coincident with the
topological transition point of the NN LKC at $u^{2}/J^{2}+v^{2}/\Delta^{2}=1$.
There should thus be a transition from a log-law entangled topological
superconducting phase to an area-law entangled trivial phase in the
NN LKC with the increase in the loss rate $v$.

\begin{figure}
	
	\includegraphics[scale=0.5]{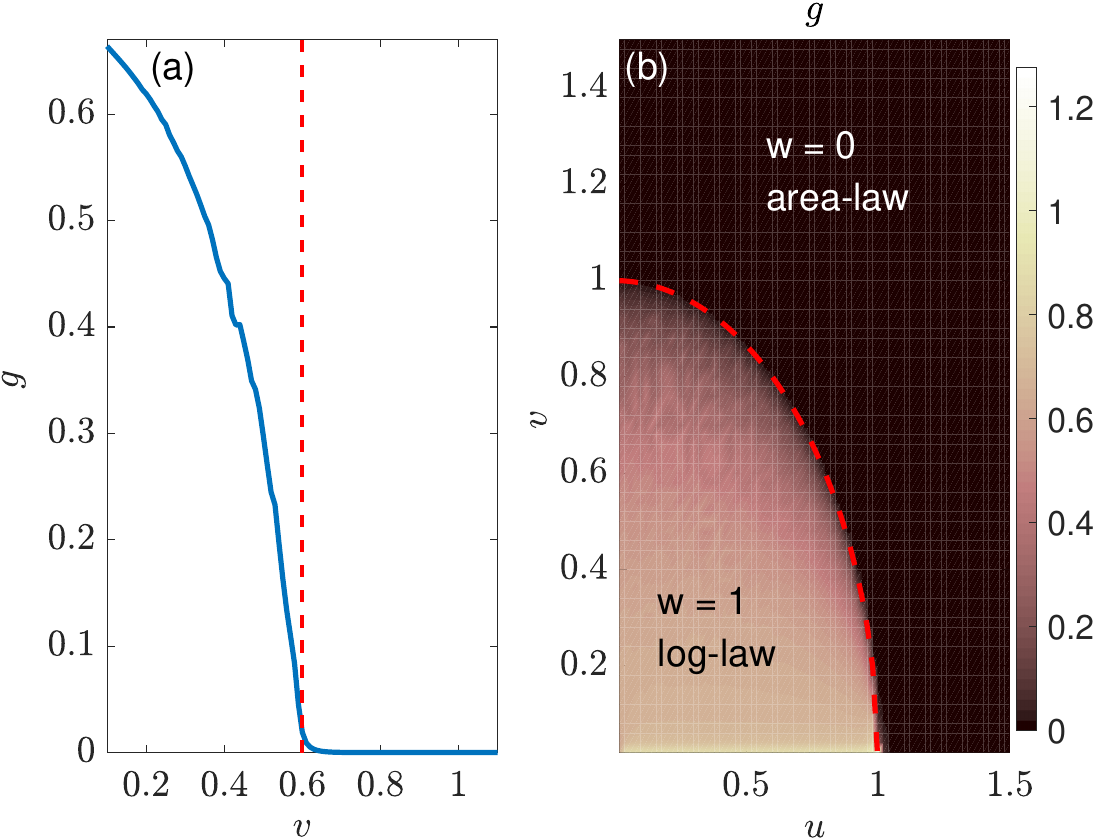}
	
	\caption{Entanglement phase transitions in the LKC with NN hopping and pairing.
		System parameters are $J=\Delta=1$ and $L=2\times10^{4}$ for both
		panels. (\textbf{a}) Gradient $g$ extracted from the data fitting $S(L,l)\sim g\ln[\sin(\pi l/L)]$
		of bipartite, steady-state EE vs. the subsystem size $l$ at different
		loss rates for $u=0.8$. The vertical dashed line highlights the phase
		transition point at $v=0.6$. (\textbf{b}) The same gradient $g$ as obtained
		in (\textbf{a}) vs. the real and imaginary parts of chemical potential $\mu=u-iv$.
		The values of $g$ at different $(u,v)$ can be read out from the
		color bar. \label{fig:LKC1EEPhsDiag}}
\end{figure}

Based on the above analysis, we could identify the entanglement phase
transition and establish the entanglement phase diagram of the NN
LKC, as reported in Figure~\ref{fig:LKC1EEPhsDiag}. To reveal the change
in scaling behaviors in the steady-state EE, we fit $S(L,l)$ with
the function $g\ln[\sin(\pi l/L)]$ in different parameter regions
and extract the coefficient $g$ as the gradient of the associated
scaling laws. In Figure~\ref{fig:LKC1EEPhsDiag}a, we take $J=\Delta=1$
and the real part of chemical potential $u=0.8$. Referring to the
topological phase boundary, $v=|\Delta|\sqrt{1-u^{2}/J^{2}}$, we
find the topological transition point to be $v=0.6$ in this case.
It is clear that we have a finite $g$ when $v<0.6$ and a vanishing
$g$ for $v>0.6$ in Figure~\ref{fig:LKC1EEPhsDiag}a, which verifies
the presence of two different entangling phases (log-law vs. area-law)
and an entanglement phase transition driven by the change in loss
rate $v$ in the NN LKC.

Finally, we present the gradient $g$ extracted from the fitting $S(L,l)\sim g\ln[\sin(\pi l/L)]$
at different $(u,v)$ in Figure~\ref{fig:LKC1EEPhsDiag}b, generating
the entanglement phase diagram of the system. We observe two different
regions with distinct entanglement features ($g>0$ vs. $g=0$), which
are separated by the boundary line $v=|\Delta|\sqrt{1-u^{2}/J^{2}}$
for $v\neq0$ and $|u|\leq|J|$. A direct comparison with the topological
phase diagram in Fig.~\ref{fig:LKC1EW}(a) leads to the conclusion
that for the NN LKC, the bipartite EE of steady states follows the
log-law scaling vs. the subsystem size in the topological phase (with
winding number $w=1$) and becomes independent of the subsystem size
in the trivial phase (with $w=0$). At $v\neq0$, a log-law to area-law
entanglement phase transition could happen following the transition
of the system from a topologically nontrivial to a trivial phase.
This entanglement transition could thus offer a unique dynamical probe
to topological phase transitions in non-Hermitian Kitaev chains.

A possible mechanism behind the loss-driven entanglement transition
in the NN LKC is as follows. With $v>0$, the particles tend to populate
the energy levels with positive imaginary parts after being evolved
over a long time duration. In the regime with $v>|\Delta|\sqrt{1-u^{2}/J^{2}}$,
such energy levels are separated apart from those with negative imaginary
parts by a dissipation gap {[}see Figure~\ref{fig:LKC1EW}c{]}. The
particle distribution of late-time steady state then mimics a 1D normal
band insulator at half filling, whose bipartite EE is expected to
follow an area-law scaling versus the system size. In the regime with
$v<|\Delta|\sqrt{1-u^{2}/J^{2}}$, there is no dissipation gap between
the energy levels with positive and negative imaginary parts {[}see
Figure~\ref{fig:LKC1EW}b{]}. After a long evolution time, the steady-state
population tends to form an effective Fermi surface at $E=0$ along
the ${\rm Re}E$ axis, which has two crossing points with the bulk spectrum.
Therefore, the particle distribution of steady state is close to a
critical metallic phase in 1D systems, whose bipartite EE is expected
to scale logarithmically with respect to the system size \cite{EEQFT01,EEQFT02,EEQFT03}.

In the next subsection, we investigate entanglement transitions in
an LKC with NNN hopping and pairing terms, whose topological phases
could possess larger winding numbers. We will see that the connections
between topological and entanglement phase transitions found in this
subsection could be extended to more general situations for non-Hermitian
topological superconductors.

\subsection{LKC with Next-Nearest-Neighbor Hopping and Pairing\label{subsec:NNNLKC}}
\begin{figure}	
	\includegraphics[scale=0.6]{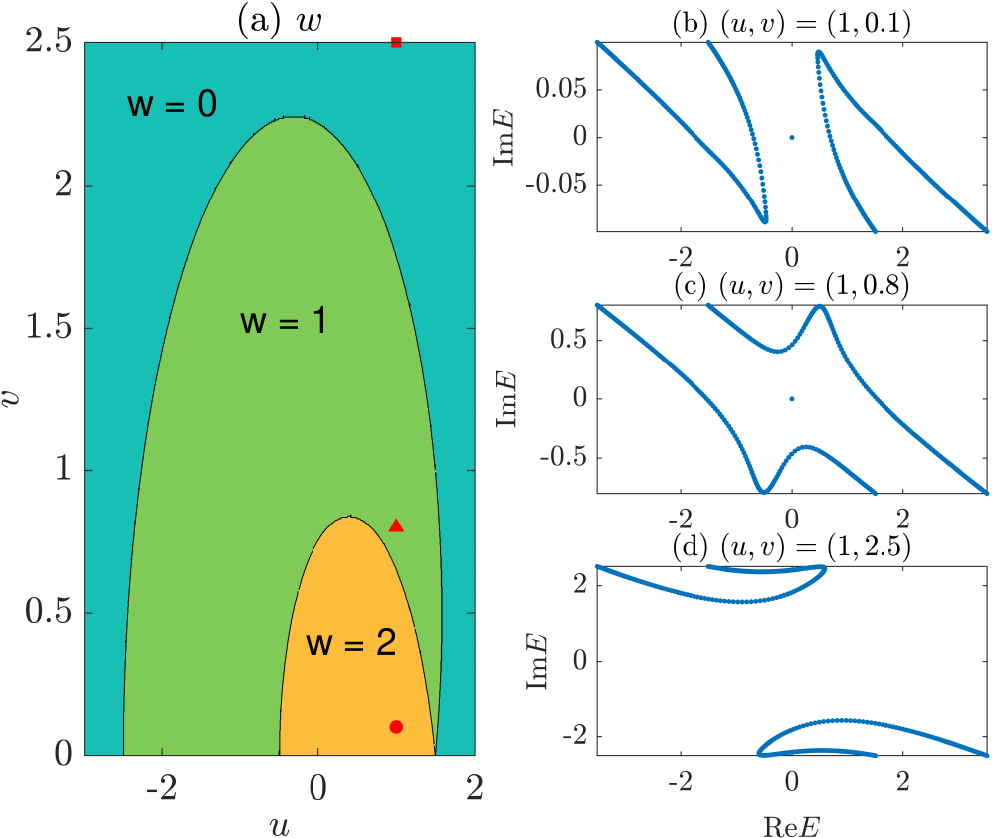}
	
	\caption{{Topological} phase diagram and typical spectra of the LKC with NNN
		hopping and pairing. (\textbf{a}) shows the winding number $w$ vs. the real
		and imaginary parts of chemical potential $u$ and $v$. The yellow,
		green and blue regions have $w=2$, $1$ and $0$, respectively. The
		red solid dot of (\textbf{a}) resides at $(u,v)=(1,0.1)$, with the associated
		spectrum of $\hat{H}_{2}$ shown in (\textbf{b}). The red solid triangle of
		(\textbf{a}) is located at $(u,v)=(1,1.1)$, and the associated spectrum of
		$\hat{H}_{2}$ is given in (\textbf{c}) on the complex energy plane. The red
		solid square of (a) lies at $(u,v)=(1,2.5)$, and the associated spectrum
		of $\hat{H}_{2}$ is shown in (\textbf{d}). Other system parameters are $J_{1}=\Delta_{1}=1$
		and $J_{2}=\Delta_{2}=1.5$ for all panels. \label{fig:LKC2EW}}
\end{figure}

We now consider an LKC with second-neighbor hopping and pairing terms.
Referring to Equation~(\ref{eq:H}), we set the coupling range $r=2$,
and the resulting Hamiltonian of the system reads
\begin{equation}
	\hat{H}_{2}=\frac{1}{2}\sum_{n}\left[\mu(2\hat{c}_{n}^{\dagger}\hat{c}_{n}-1)+\sum_{r=1,2}(J_{r}\hat{c}_{n}^{\dagger}\hat{c}_{n+r}+\Delta_{r}\hat{c}_{n}\hat{c}_{n+r}+{\rm H.c.})\right].\label{eq:H2}
\end{equation}
The non-Hermitian effect is again introduced by setting the chemical
potential $\mu=u-iv$ with $v>0$. In momentum space, it can be shown
that there are four possible critical quasimomenta $\pm k_{0}^{\pm}=\pm\arccos\{[-J_{1}\pm\sqrt{J_{1}^{2}+8J_{2}(J_{2}-u)}]/(4J_{2})\}$,
where the two bulk energy bands of $\hat{H}_{2}$ could touch with
each other at $E=0$ \cite{NHTSC07}. The topological phase boundaries
of the system in parameter space can then be obtained by solving the
equations $\sin(k_{0}^{\pm})(\Delta_{1}+2\Delta_{2}\cos k_{0}^{\pm})=\pm v$
\cite{NHTSC07}. In Figure~\ref{fig:LKC2EW}a, we plot these phase
boundaries as black lines in the $u-v$ plane and compute the topological
winding number $w$ {[}Equation~(\ref{eq:w}){]} in different parameter
regions. The results show that there are three different topological
phases with winding numbers $w=0,1,2$. The presence of NNN hopping
and coupling terms allows us to obtain a non-Hermitian topological superconducting
phase with a larger winding number $w=2$. 

In the topological phase with $w=2$, the bulk energy spectrum of
the system is gapped along the ${\rm Re}E$ axis, and there are two
pairs of Majorana edge modes at $E=0$ under the OBC {[}see Figure~\ref{fig:LKC2EW}b{]}.
With the increase in $v$, the system could undergo a phase transition
through level crossings at $E=0$, after which it enters another topological
phase with $w=1$. In this phase, the bulk spectrum of the system
holds a line energy gap along a certain angle $\theta\in(0,\pi/2)$
with respect to the ${\rm Re}E$ axis, and a single pair of Majorana
edge modes can be found inside the gap at $E=0$ under the OBC {[}see
Figure~\ref{fig:LKC2EW}c{]}. Note in passing that this non-Hermitian
topological superconducting phase is different from the $w=1$ phase
of NN LKC {[}see Figure~\ref{fig:LKC1EW}{]}, whose spectral gap lies
instead along the ${\rm Re}E$ axis. With the further increase in
$v$, the system would experience a second phase transition and finally
entering a trivial phase with $w=0$. In this phase, the spectrum
of the system develops a line gap along the ${\rm Im}E$ axis, and
there are no zero-energy Majorana edge modes within the gap under
the OBC {[}see Figure~\ref{fig:LKC2EW}d{]}.

\begin{figure}
	
	\includegraphics[scale=0.6]{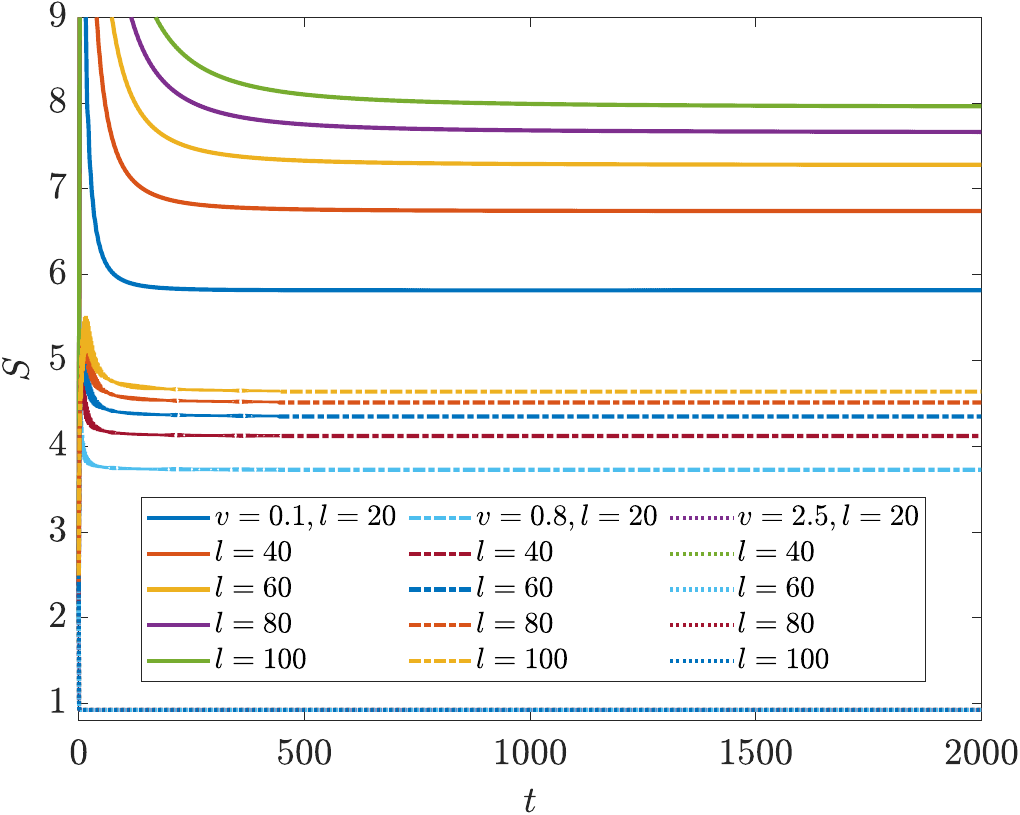}
	
	\caption{{Bipartite} EE vs. time $t$ for the LKC with NNN hopping and pairing
		for the loss rate $v=0.1$ (solid lines), $0.8$ (dash-dotted lines)
		and $2.5$ (dotted lines). Other system parameters are $J_{1}=\Delta_{1}=1$,
		$J_{2}=\Delta_{2}=1.5$ and $u=1$. $l$ denotes the subsystem size
		and the total lattice size is $L=2\times10^{4}$. \label{fig:LKC2EEvst}}
\end{figure}

After unveiling the rich spectral and topological features of the
NNN LKC, we are ready to explore its entanglement dynamics. Referring
again to Equations (\ref{eq:Psit})--(\ref{eq:St}), we can find the
evolution of bipartite EE $S(t)$ of the NNN LKC with a large system
size $L$ under PBC. Exemplary results vs. different sizes $l$ of
the subsystem A and different loss rates are presented in Figure~\ref{fig:LKC2EEvst}.
We have also considered a half-filled system and used Equation~(\ref{eq:psik0})
as the initial state at different quasimomenta throughout the calculations.
The results show that in the case with a large loss rate ($v=2.5$
in Figure~\ref{fig:LKC2EEvst}), the EE does not change with the increase
in subsystem size $l$ after a long evolution time (dotted lines in
Figure~\ref{fig:LKC2EEvst}), which implies an area-law scaling of the
steady-state EE vs. $l$. This case corresponds to the spectrum of
NNN LKC in Figure~\ref{fig:LKC2EW}d, where there exists a dissipation
gap, and the system resides in a topologically trivial phase. In the
situation with a small loss rate ($v=0.1$ in Figure~\ref{fig:LKC2EEvst}),
the EE grows monotonically with the increase in the subsystem size
$l$ after a long-time evolution (solid lines in Figure~\ref{fig:LKC2EEvst}).
It corresponds to the case with a line energy gap along the ${\rm Re}E$
axis in the spectrum {[}Figure~\ref{fig:LKC2EW}b{]}, and the system
resides in a topological phase with winding number $w=2$. When the
loss rate takes an intermediate value ($v=0.8$ in Figure~\ref{fig:LKC2EEvst}),
the EE again rises with the increase in the subsystem size $l$ in
the late time regime. Nevertheless, its growth rate vs. $l$ is smaller
compared to the case of $v=0.1$. The system also belongs to a topological
phase with a smaller winding number $w=1$ and possessing a gapped
spectrum as shown in Figure~\ref{fig:LKC2EW}c. These observations
suggest that for the NNN LKC, the bipartite EE in steady states may
possess different scaling properties when the system belongs to distinct
topological superconducting phases with distinguishable spectral features.

\begin{figure}
	
	\includegraphics[scale=0.5]{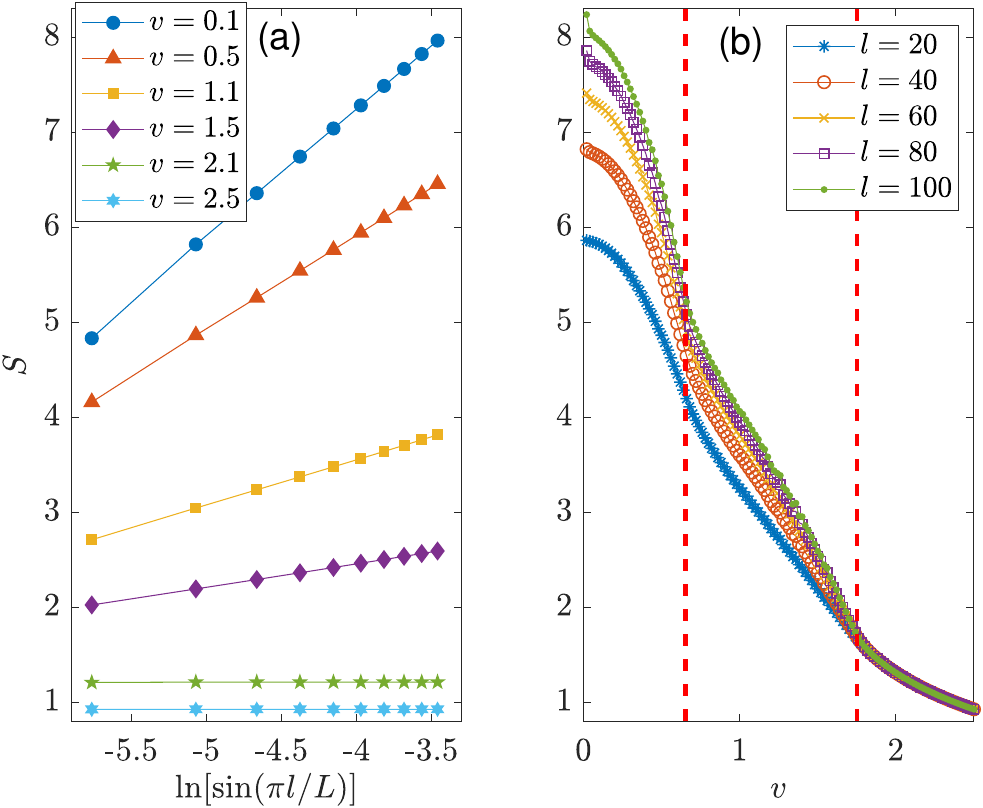}
	
	\caption{{Bipartite}  EE of steady states vs. (\textbf{a}) the subsystem size $l$, and
		(\textbf{b}) the loss rate $v$ for the LKC with NNN hopping and pairing. Other
		system parameters are $J_{1}=\Delta_{1}=1$, $J_{2}=\Delta_{2}=1.5$
		and $u=1$ for both panels. The lattice size of the whole system is
		$L=2\times10^{4}$. Vertical dashed lines in (\textbf{b}) denote two topological
		transition points of the system. \label{fig:LKC2EEvslv}}
\end{figure}

To further decode the scaling laws of bipartite EE in steady states,
we consider the evolution of the system over a long time duration $T=2000$ and
obtain the final values of EE $S(L,l)$ at different subsystem sizes
$l$ and loss rates, as shown in Figure~\ref{fig:LKC2EEvslv}. The methodology
of computing $S(L,l)$ here is in parallel with that employed in 
Section~\ref{subsec:NNLKC}. We find that in the weak and intermediate dissipation
regions {[}$v=0.1,0.5$ and $v=1.1,1.5$ in Figure~\ref{fig:LKC2EEvslv}a{]},
the EE $S(L,l)$ is proportional to $\ln[\sin(\pi l/L)]$ for a fixed
system size $L$. In the limit of large $L$, this relation reduces
to the log-law scaling $S\sim\ln l$. In the strong dissipation region
{[}$v=2.1,2.5$ in Figure~\ref{fig:LKC2EEvslv}a{]}, the steady-state
EE becomes independent of $l$, leading to an area-law entangled phase
with $S\sim l^{0}$. More precisely, under the condition $v>0$, we
numerically find the following three parameter regions in which the
bipartite EE of steady states show distinguishable scaling behaviors
in the limit $L\rightarrow\infty$, i.e., 
\begin{equation}
	S(L,l)\sim\begin{cases}
		g\ln l, & v<\min|h_{y}(k_{0}^{\pm})|,\\
		g'\ln l, & \min|h_{y}(k_{0}^{\pm})|<v<\max|h_{y}(k_{0}^{\pm})|,\\
		l^{0}, & v>\max|h_{y}(k_{0}^{\pm})|.
	\end{cases}\label{eq:EE2vsl}
\end{equation}
Here, the coefficients $g$ and $g'$ both depend on the system parameters,
and we always have $g>g'$. The expression of $h_{y}(k)$ is given
by Equation~(\ref{eq:hyk}) after setting $r=1,2$.  $k_{0}^{\pm}\in[-\pi,\pi)$
are the critical quasimomenta where the energy bands could touch when
a topological phase transition happens \cite{NHTSC07}. The three distinct
scaling regions of $S(L,l)$ are clearly visible from the EE vs. loss
rate $v$ in Figure~\ref{fig:LKC2EEvslv}b. Notably, the separation
points between these entangling phases tend to be identical to
the topological transition points of the NNN LKC \cite{NHTSC07}.
Despite a transition from a log-law entangled topological phase to
an area-law entangled trivial phase, there should also be a possible
entanglement transition between two non-Hermitian topological superconducting
phases with winding numbers $w=2$ and $w=1$ following the variation
in the loss rate $v$.

\begin{figure}
	
	\includegraphics[scale=0.5]{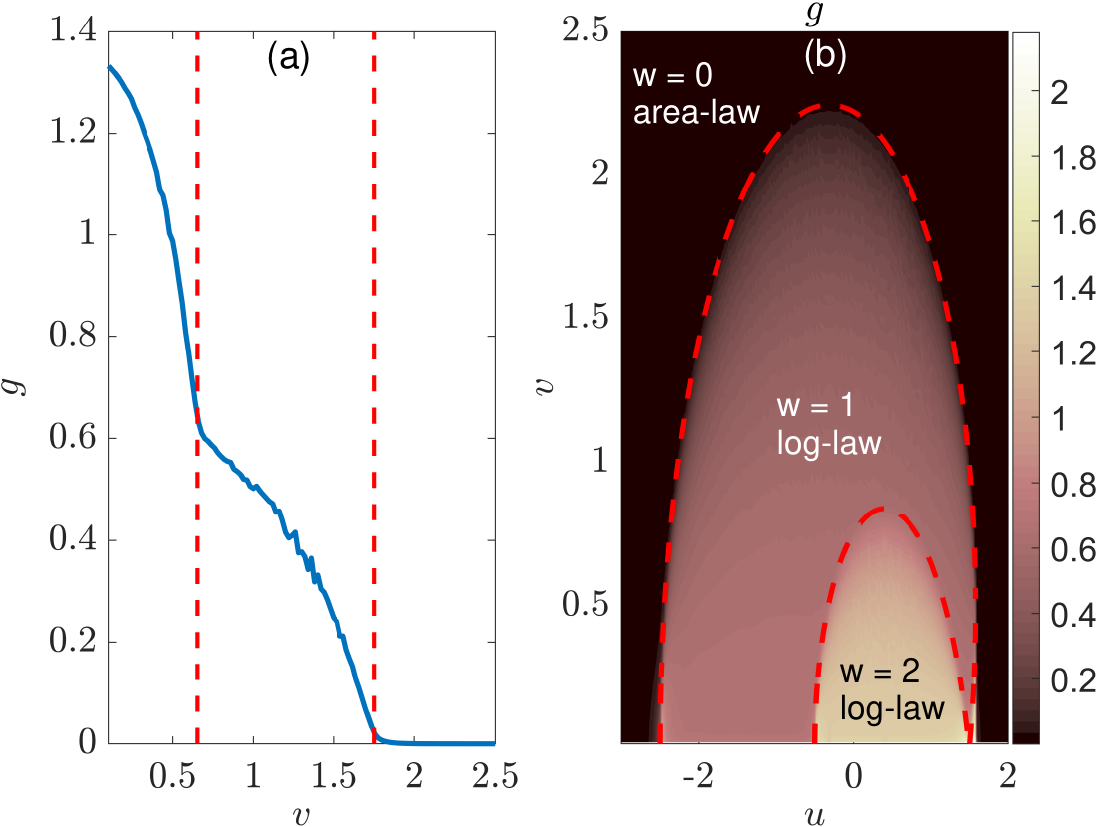}
	
	\caption{{Entanglement} phase transitions in the LKC with NNN hopping and pairing.
		System parameters are $J_{1}=\Delta_{1}=1$, $J_{2}=\Delta_{2}=1.5$
		and $L=2\times10^{4}$ for both panels. (\textbf{a}) Gradient $g$ extracted
		from the data fitting $S(L,l)\sim g\ln[\sin(\pi l/L)]$ of bipartite,
		steady-state EE vs. the subsystem size $l$ at different loss rate
		$v$ for $u=1$. Vertical dashed lines highlight topological transition
		points of the system. (\textbf{b}) The same gradient $g$ as obtained in (\textbf{a})
		vs. the real and imaginary parts of chemical potential $\mu=u-iv$.
		The values of $g$ at different $(u,v)$ can be figured out from the
		color bar. \label{fig:LKC2EEPhsDiag}}
\end{figure}

To confirm the presence of entanglement transitions in the NNN LKC
and build their connections with topological phase transitions, we
perform numerical fitting for the bipartite EE of steady states as
$S(L,l)\sim g\ln[\sin(\pi l/L)]$ and extract the coefficients $g$
at different system parameters. The results are shown in Figure~\ref{fig:LKC2EEPhsDiag}.
The two red dashed lines in Figure~\ref{fig:LKC2EEPhsDiag}a correspond
to the loss rates where topological phase transitions happen in the
system \cite{NHTSC07}. We see that they separate the configuration
of $g$ into three distinguishable regions. In the left and middle
regions (with weak and intermediate dissipation), the values of $g$
are finite and decrease gradually with the increase in $v$. The system
then belongs to log-law entangled phases in these domains. In the
right region, the gradient $g$ becomes pinned to zero, which implies
that the system has entered an area-law entangled phase with $S(L,l)\sim l^{0}$.
It is noteworthy that the derivatives of $g$ with respect to the
loss rate $v$ undergo two discontinuous changes at the topological
phase transition points. Therefore, the loss-induced topological transitions
in the NNN LKC also accompany entanglement phase transitions characterized
by different scaling laws in the bipartite EE of steady states.

Finally, we present the scaling coefficients $g$ of $S(L,l)$ vs.
the real and imaginary parts of chemical potential {$\mu$} 
in 
Figure~\ref{fig:LKC2EEPhsDiag}b, which forms the entanglement phase diagram
of the system. In comparison with the topological phase diagram in
Figure~\ref{fig:LKC2EW}a, we conclude that the topological phases
of NNN LKC with different winding numbers $w=2$, $1$ and $0$ indeed
exhibit different scaling laws in the EE vs. system sizes. Both the
two topologically nontrivial phases are log-law entangled, whereas
the topologically trivial phase is area-law entangled. Each topological
phase transition further goes hand-in-hand with an entanglement phase~transition. 

The physical mechanism behind these loss-driven entanglement transitions
is similar to that discussed for the NN LKC in Section~\ref{subsec:NNLKC}.
Starting with an initial state at half-filling, the final population
of the steady state tends to fill the energy levels with positive
imaginary parts after a long evolution time. The interface between
filled and empty states then possesses four, two and no crossing points
with the bulk energy spectrum of $\hat{H}_{2}$ in the topological
phases with winding numbers $w=2$, $1$ and $0$, respectively, (see
Figure~\ref{fig:LKC2EW}), yielding two critical-type, log-law entangled
phases and a gapped, area-law entangled phase. The entanglement phase
transitions identified here may also provide us with an alternative
strategy to dynamically probe and distinguish non-Hermitian topological
superconducting phases with large winding numbers.

\section{Conclusions\label{sec:Sum}}

In this work, we unveiled entanglement phase transitions in 1D
non-Hermitian topological superconductors. Our investigation focused
on Kitaev chains exhibiting onsite particle losses, revealing distinct
scaling laws of EE corresponding to different non-Hermitian superconducting
phases. Specifically, the bipartite EE of steady states demonstrated
an area-law scaling in the topologically trivial phase and a log-law
scaling in the nontrivial phase. Notably, the scaling coefficients
of the log-law become distinguishable when the system is situated
in superconducting phases with distinct topological winding numbers.
Furthermore, we identified entanglement phase transitions coinciding
with topological phase transitions in non-Hermitian Kitaev chains
at identical system parameters. Our study established generic connections
among spectral, topological, and entanglement phase transitions in
two distinct lossy Kitaev chains with varying hopping and pairing
ranges. These findings not only uncovered the richness and unique
characteristics of entanglement transitions in a class of non-Hermitian
systems, but also introduced an efficient dynamical probe for detecting
and distinguishing different non-Hermitian superconducting phases
with diverse spectral and topological properties.

In future research, exploring entanglement phase transitions in non-Hermitian
topological superconductors with disorder, subject to time-periodic
driving, and extending the investigation to higher spatial dimensions
holds considerable interest. The critical properties of entanglement
transitions in non-Hermitian systems and their responses to many-body
interactions deserve more thorough explorations. Additionally, the
experimental realization of non-Hermitian Kitaev chains and the detection
of entanglement and topological phase transitions within these systems
offer intriguing directions for future studies.

\begin{acknowledgments}
	This work is supported by the National Natural Science Foundation of China (Grants No.~12275260, No.~12047503 and No.~11905211), the Fundamental Research Funds for the Central Universities (Grant No.~202364008), and the Young Talents Project of Ocean University of China.
\end{acknowledgments}

\appendix

\section{Derivation of the Non-Hermitian Hamiltonian}\label{sec:AppA}

In this Appendix, we obtain our LKC Hamiltonian in Equation~(\ref{eq:H}) following the quantum trajectory approach \cite{NHEPT05}. We consider a Markovian open quantum system described by the Lindblad master Equation ($\hbar=1$)
\begin{equation}
	\frac{d}{dt}{\hat \rho}=-i[{\hat H}_0,{\hat \rho}]+\sum_n \left( {\hat L}_n{\hat \rho}{\hat L}_n^{\dagger}-\frac{1}{2}\{ {\hat L}_n^{\dagger}{\hat L}_n,{\hat \rho}\} \right).\label{eq:ME1}
\end{equation}
Here, ${\hat \rho}$ is the density matrix of the system. ${\hat H}_0={\hat H}^{\dagger}_0$ is the Hermitian Hamiltonian that induces the coherent dynamics. $\{{\hat L}_n\}$ is the jump operator depicting the coupling of the system to the external environment. A re-organization of the terms in the master equation yields
\begin{equation}
	\frac{d}{dt}{\hat \rho}=-i[{\hat H}{\hat \rho}-{\hat \rho}{\hat H}^{\dagger}]+\sum_n {\hat L}_n{\hat \rho}{\hat L}_n^{\dagger},\label{eq:ME2}
\end{equation}
where the effective Hamiltonian is
\begin{equation}
	{\hat H}\equiv{\hat H}_0-\frac{i}{2}\sum_n {\hat L}_n^{\dagger}{\hat L}_n.\label{eq:Heff}
\end{equation}
The last term in Equation~(\ref{eq:ME2}) characterizes each quantum trajectory under stochastic loss events. $\sqrt{dt}{\hat L}_n$ might be regarded as a measurement operator for a signal $n$ in the time interval $[t,t+dt]$. $1-i{\hat H}dt$ may be further viewed as a measurement operator for zero signals. Under continuous monitoring and post-selection of null measurement outcomes, the quantum jump terms $\sum_n {\hat L}_n{\hat \rho}{\hat L}_n^{\dagger}$ are ignored (no-click limit). The dynamics are then described by the effective non-Hermitian Hamiltonian in Equation~(\ref{eq:Heff}).

To arrive at our LKC Hamiltonian in Equation~(\ref{eq:H}), we set the ${\hat H}_0$ and ${\hat L}_n$ in Equation (\ref{eq:Heff}) to be
\begin{equation}
	\hat{H}_0=\frac{1}{2}\sum_{n}\left[u(2\hat{c}_{n}^{\dagger}\hat{c}_{n}-1)+\sum_{r}(J_{r}\hat{c}_{n}^{\dagger}\hat{c}_{n+r}+\Delta_{r}\hat{c}_{n}\hat{c}_{n+r}+{\rm H.c.})\right]\label{eq:H0}
\end{equation}
and
\begin{equation}
	{\hat L}_n=\sqrt{2v}{\hat c}_n.\label{eq:Ln}
\end{equation}
Inserting them into Equation~(\ref{eq:Heff}) gives us the LKC Hamiltonian in Equation~(\ref{eq:H}) (up to a global constant term that does not affect the wavepacket dynamics), where the onsite potential amplitude $\mu=u-iv$.

\section{Momentum-Space Generator of Correlation Matrix}\label{sec:AppB}

In this Appendix, we discuss how to obtain the $k$-space generator
{[}Equation~(\ref{eq:Ckt}){]} of the single-particle correlation matrix in
real space. We first transform the Hamiltonian in Equation~(\ref{eq:H})
to another real-space fermionic basis through the Bogoliubov-de Gennes
(BdG) transformation \cite{BdG}. Let us introduce creation and annihilation
operators of normal fermions $\hat{f}_{j}^{\dagger}$ and $\hat{f}_{j}$,
which satisfy the anticommutation relations
\begin{equation}
	\{\hat{f}_{i},\hat{f}_{j}\}=\{\hat{f}_{i}^{\dagger},\hat{f}_{j}^{\dagger}\}=0,\qquad\{\hat{f}_{i},\hat{f}_{j}^{\dagger}\}=\delta_{ij}.\label{eq:fj}
\end{equation}
With these operators, we can expand the annihilation operator $\hat{c}_{n}$
in Equation~(\ref{eq:Ckt}) as
\begin{equation}
	\hat{c}_{n}=\sum_{j=1}^{L}(\varphi_{jn}\hat{f}_{j}+\phi_{jn}\hat{f}_{j}^{\dagger}).\label{eq:cnfj}
\end{equation}
Here, the coefficients $\varphi_{jn}$ and $\phi_{jn}$ satisfy
the normalization condition $\sum_{j=1}^{L}(\varphi_{jn}^{2}+\phi_{jn}^{2})=1$.
Assuming that the Hamiltonian $\hat{H}$ in Equation~(\ref{eq:H}) is diagonalized
on the basis\linebreak   $\{\hat{f}_{1},\ldots,\hat{f}_{L},\hat{f}_{1}^{\dagger},\ldots,\hat{f}_{L}^{\dagger}\}$,
we can express $\hat{H}$ as
\begin{equation}
	\hat{H}=\sum_{\ell}E_{\ell}\hat{f}_{\ell}^{\dagger}\hat{f}_{\ell}.\label{eq:HH}
\end{equation}
Using the commutator formula $[AB,C]=A\{B,C\}-\{A,C\}B$ together
with Equations (\ref{eq:cnfj}) and (\ref{eq:HH}), we find that
\begin{equation}
	[\hat{H},\hat{f}_{j}]=-E_{j}\hat{f}_{j},\qquad[\hat{H},\hat{f}_{j}^{\dagger}]=E_{j}\hat{f}_{j}^{\dagger},\label{eq:Hfj}
\end{equation}
\begin{equation}
	[\hat{H},\hat{c}_{n}]=-\sum_{j}E_{j}(\varphi_{jn}\hat{f}_{j}-\phi_{jn}\hat{f}_{j}^{\dagger}).\label{eq:Hcn}
\end{equation}
Meanwhile, from Equations~(\ref{eq:H}) and (\ref{eq:cnfj}), we can equivalently
express the commutator $[\hat{H},\hat{c}_{n}]$ as
\begin{alignat}{1}
	[\hat{H},\hat{c}_{n}]= & -\sum_{j}\mu(\varphi_{jn}\hat{f}_{j}+\phi_{jn}\hat{f}_{j}^{\dagger})\nonumber \\
	& -\sum_{j}\sum_{r}\frac{J_{r}}{2}(\varphi_{jn+r}+\varphi_{jn-r})\hat{f}_{j}\nonumber \\
	& -\sum_{j}\sum_{r}\frac{J_{r}}{2}(\phi_{jn+r}+\phi_{jn-r})\hat{f}_{j}^{\dagger}\nonumber \\
	& +\sum_{j}\sum_{r}\frac{\Delta_{r}}{2}(\phi_{jn+r}-\phi_{jn-r})\hat{f}_{j}\nonumber \\
	& +\sum_{j}\sum_{r}\frac{\Delta_{r}}{2}(\varphi_{jn+r}-\varphi_{jn-r})\hat{f}_{j}^{\dagger}.\label{eq:Hcn2}
\end{alignat}
Equationating the right hand sides of Equations~(\ref{eq:Hcn}) and (\ref{eq:Hcn2})
and dropping the redundant index $j$, we arrive at the following
BdG self-consistent equations
\begin{equation}
	+\mu\varphi_{n}+\sum_{r}\frac{J_{r}}{2}(\varphi_{n-r}+\varphi_{n+r})+\sum_{r}\frac{\Delta_{r}}{2}(\phi_{n-r}-\phi_{n+r})=E\varphi_{n},\label{eq:BdG1}
\end{equation}
\begin{equation}
	-\mu\phi_{n}-\frac{J_{r}}{2}\sum_{r}(\phi_{n-r}+\phi_{n+r})-\sum_{r}\frac{\Delta_{r}}{2}(\varphi_{n-r}-\varphi_{n+r})=E\phi_{n}.\label{eq:BdG2}
\end{equation}

Next, it can be verified that Equations (\ref{eq:BdG1}) and (\ref{eq:BdG2})
can be generated by a fermionic Hamiltonian $\hat{{\cal H}}$. It
owns two internal degrees of freedom $(\alpha,\beta)$ in each unit
cell but has no superconducting pairing terms, {i.e.,} 

\begin{equation}
	\hat{{\cal H}}=\frac{1}{2}\sum_{n}\left[\mu\hat{{\bf c}}_{n}^{\dagger}\sigma_{z}\hat{{\bf c}}_{n}+\sum_{r}\hat{{\bf c}}_{n}^{\dagger}(J_{r}\sigma_{z}-i\Delta_{r}\sigma_{y})\hat{{\bf c}}_{n+r}+{\rm H.c.}\right],\label{eq:HHH}
\end{equation}
where $\hat{{\bf c}}_{n}^{\dagger}=(\hat{c}_{n,\alpha}^{\dagger},\hat{c}_{n,\beta}^{\dagger})$.
Taking the PBC and applying the Fourier transformation $\hat{{\bf c}}_{n}^{\dagger}=\frac{1}{\sqrt{L}}\sum_{k}e^{-ikn}\hat{{\bf c}}_{k}^{\dagger}$
with the momentum-space creation operator $\hat{{\bf c}}_{k}^{\dagger}=(\hat{c}_{k,\alpha}^{\dagger},\hat{c}_{k,\beta}^{\dagger})$,
we arrive at $\hat{{\cal H}}=\sum_{k}\hat{{\bf c}}_{k}^{\dagger}{\cal H}(k)\hat{{\bf c}}_{k}$.
Here, the Bloch Hamiltonian is 
\begin{equation}
	{\cal H}(k)=\sum_{r}\Delta_{r}\sin(kr)\sigma_{y}+[\mu+\sum_{r}J_{r}\cos(kr)]\sigma_{z},\label{eq:HHHk}
\end{equation}
which is formally identical to the $H(k)$ in Equation~(\ref{eq:Hk}).
In the basis of $\hat{{\bf c}}_{k}$ and $\hat{{\bf c}}_{k}^{\dagger}$,
we can now compute the $k$-space generator $C_{k}(t)$ of correlation
matrix in Equation~(\ref{eq:Clmt}) as \cite{NHEPT06}
\begin{equation}
	C_{k}(t)=\begin{pmatrix}\langle\hat{c}_{k,\alpha}^{\dagger}\hat{c}_{k,\alpha}\rangle_{t} & \langle\hat{c}_{k,\alpha}^{\dagger}\hat{c}_{k,\beta}\rangle_{t}\\
		\langle\hat{c}_{k,\beta}^{\dagger}\hat{c}_{k,\alpha}\rangle_{t} & \langle\hat{c}_{k,\beta}^{\dagger}\hat{c}_{k,\beta}\rangle_{t}
	\end{pmatrix}.\label{eq:Ckt2}
\end{equation}
Since the system is noninteracting, we can express the single particle
operators in the first quantized formalism as
\begin{equation}
	\hat{c}_{k,\alpha}^{\dagger}\hat{c}_{k,\alpha}\rightarrow|\alpha\rangle\langle\alpha|,\qquad\hat{c}_{k,\beta}^{\dagger}\hat{c}_{k,\beta}\rightarrow|\beta\rangle\langle\beta|,
\end{equation}
\begin{equation}
	\hat{c}_{k,\alpha}^{\dagger}\hat{c}_{k,\beta}\rightarrow|\alpha\rangle\langle\beta|,\qquad\hat{c}_{k,\beta}^{\dagger}\hat{c}_{k,\alpha}\rightarrow|\beta\rangle\langle\alpha|,
\end{equation}
where $\{|\alpha\rangle,|\beta\rangle\}$ forms the two by two internal
subspace of the Bloch Hamiltonian ${\cal H}(k)$ with the identity matrix
$\sigma_{0}\equiv|\alpha\rangle\langle\alpha|+|\beta\rangle\langle\beta|$.
Finally, with the help of Pauli matrices, we can identify
\begin{equation}
	\sigma_{x}=|\alpha\rangle\langle\beta|+|\beta\rangle\langle\alpha|,\label{eq:sx}
\end{equation}
\begin{equation}
	\sigma_{y}=-i(|\alpha\rangle\langle\beta|-|\beta\rangle\langle\alpha|),\label{eq:sy}
\end{equation}
\begin{equation}
	\sigma_{z}=|\alpha\rangle\langle\alpha|-|\beta\rangle\langle\beta|.\label{eq:sz}
\end{equation}
Therefore, the matrix elements of $C_{k}(t)$ in Equation~(\ref{eq:Ckt2})
can be equivalently expressed as $\langle\hat{c}_{k,\alpha}^{\dagger}\hat{c}_{k,\alpha}\rangle_{t}=\langle\sigma_{0}+\sigma_{z}\rangle_{t}/2$,
$\langle\hat{c}_{k,\beta}^{\dagger}\hat{c}_{k,\beta}\rangle_{t}=\langle\sigma_{0}-\sigma_{z}\rangle_{t}/2$,
$\langle\hat{c}_{k,\alpha}^{\dagger}\hat{c}_{k,\beta}\rangle_{t}=\langle\sigma_{x}+i\sigma_{y}\rangle_{t}/2$
and $\langle\hat{c}_{k,\beta}^{\dagger}\hat{c}_{k,\alpha}\rangle_{t}=\langle\sigma_{x}-i\sigma_{y}\rangle_{t}/2$,
yielding the expression of $C_{k}(t)$ in Equation~(\ref{eq:Ckt}).

\end{document}